\documentclass[aps,amsmath,amssymb,floatfix,superscriptaddress,showpacs,nofootinbib]{revtex4}
\usepackage{graphicx}

\newcommand{\id}{{\mathbf 1}}

\newcommand{\cF}{\mathcal{F}}
\newcommand{\cS}{\mathcal{S}}
\newcommand{\bm}{{\bf m}}
\newcommand{\bp}{{\bf p}}
\newcommand{\br}{{\bf r}}
\newcommand{\bt}{{\bf t}}

\DeclareMathOperator{\sign}{sign}
\DeclareMathOperator*{\hf}{Hf}
\DeclareMathOperator*{\pf}{Pf}
\allowdisplaybreaks[4]

\font\numbers=cmss12
\font\upright=cmu10 scaled\magstep1
\def\stroke{\vrule height8pt width0.4pt depth-0.1pt}
\def\topfleck{\vrule height8pt width0.5pt depth-5.9pt}
\def\botfleck{\vrule height2pt width0.5pt depth0.1pt}
\def\Zmath{\vcenter{\hbox{\numbers\rlap{\rlap{Z}\kern
0.8pt\topfleck}\kern 2.2pt
                   \rlap Z\kern 6pt\botfleck\kern 1pt}}}
\def\Qmath{\vcenter{\hbox{\upright\rlap{\rlap{Q}\kern
                   3.8pt\stroke}\phantom{Q}}}}
\def\Nmath{\vcenter{\hbox{\upright\rlap{I}\kern 1.7pt N}}}
\def\Cmath{\vcenter{\hbox{\upright\rlap{\rlap{C}\kern
                   3.8pt\stroke}\phantom{C}}}}
\def\Rmath{\vcenter{\hbox{\upright\rlap{I}\kern 1.7pt R}}}
\def\Z{\ifmmode\Zmath\else$\Zmath$\fi}
\def\Q{\ifmmode\Qmath\else$\Qmath$\fi}
\def\N{\ifmmode\Nmath\else$\Nmath$\fi}
\def\C{\ifmmode\Cmath\else$\Cmath$\fi}
\def\R{\ifmmode\Rmath\else$\Rmath$\fi}

\def\bar{\overline}

\def\*{\star}
\def\[{\left[}
\def\]{\right]}
\def\({\left(}      
\def\){\right)}

\def\2pi{\hbox{$2\pi i$}}

\def\dsl{\raise.15ex\hbox{/}\kern-.57em\partial}
\def\Dsl{\,\raise.15ex\hbox{/}\mkern-.13.5mu D}
\def\ep{\epsilon}

\def\sign{{\rm sign}}

\def\2pi{\hbox{$2\pi i$}}

\def\dsl{\raise.15ex\hbox{/}\kern-.57em\partial}
\def\Dsl{\,\raise.15ex\hbox{/}\mkern-.13.5mu D}
\font\numbers=cmss12
\font\upright=cmu10 scaled\magstep1
\def\stroke{\vrule height8pt width0.4pt depth-0.1pt}
\def\topfleck{\vrule height8pt width0.5pt depth-5.9pt}
\def\botfleck{\vrule height2pt width0.5pt depth0.1pt}
\def\Zmath{\vcenter{\hbox{\numbers\rlap{\rlap{Z}\kern
0.8pt\topfleck}\kern 2.2pt
                   \rlap Z\kern 6pt\botfleck\kern 1pt}}}
\def\Qmath{\vcenter{\hbox{\upright\rlap{\rlap{Q}\kern
                   3.8pt\stroke}\phantom{Q}}}}
\def\Nmath{\vcenter{\hbox{\upright\rlap{I}\kern 1.7pt N}}}
\def\Cmath{\vcenter{\hbox{\upright\rlap{\rlap{C}\kern
                   3.8pt\stroke}\phantom{C}}}}
\def\Rmath{\vcenter{\hbox{\upright\rlap{I}\kern 1.7pt R}}}
\def\Z{\ifmmode\Zmath\else$\Zmath$\fi}
\def\Q{\ifmmode\Qmath\else$\Qmath$\fi}
\def\N{\ifmmode\Nmath\else$\Nmath$\fi}
\def\C{\ifmmode\Cmath\else$\Cmath$\fi}
\def\R{\ifmmode\Rmath\else$\Rmath$\fi}

\def\barray{\begin{eqnarray}}
\def\earray{\end{eqnarray}}
\def\beq{\begin{equation}}
\def\eeq{\end{equation}}

\def\s{\sigma}

\def\bm{{\bf m}}

\begin{document}

\title{Chiral correlators of the Ising conformal field theory}

\author{Eddy Ardonne}
\affiliation{Nordita, Roslagstullsbacken 23, SE-106 91 Stockholm, Sweden}
\author{Germ\'an Sierra}
\affiliation{Instituto de F\'isica T\'eorica, UAM-CSIC, Madrid, Spain}
\date{\today}

\begin{abstract}
We derive explicit expressions for the conformal blocks of the Ising conformal field
theory, for the correlators of an arbitrary number of primary fields. These results are
obtained from the bosonized description of the Ising model. Interestingly, correlators
involving Majorana fermions, can be obtained in two different ways, giving rise to
identities between the `bosonic' and `fermionic' description of these correlators.
These identities are generalizations of the famous Cauchy-identity. The conformal
blocks of the Ising model are used to derive expression for the conformal blocks
of the $su(2)_2$ WZW conformal field theory.
\end{abstract}



\maketitle

\section{Introduction}

To completely specify a conformal field theory, one has to provide some data, including the central charge, the scaling dimensions of the primary fields, their fusion rules and such. However, from this data, it is far from trivial to calculate the arbitrary correlation functions of (primary) fields. In general, it is a daunting task to obtain correlation functions of more than four fields. Of course, in the case of free fields (such as a chiral boson or a Majorana fermion), one can use Wick's theorem to calculate correlation functions of an arbitrary number of fields. In almost all other cases, this seems to be utterly impossible. A notable exception is the spin field $\sigma$ (with scaling dimension $h_\sigma = 1/16$), present in the (chiral) Ising conformal field theory. This theory can be bosonized, which allows one to calculate correlation function of an arbitrary (even) number of $\sigma$ fields, and express it in a completely explicit form.

The Ising conformal field theory is relevant for various systems, which in recent years
have attracted a lot of attention in condensed matter physics. Apart from the
fractional quantum Hall effect, which we will introduce shortly, there are several
examples. Notable are the $p+ip$-type superconductors, whose vortices
have Majorana-zero modes \cite{rg00}. Kitaev introduced the `Majorana wire' \cite{kitaev00},
a one-dimensional wire, which harbors Majorana fermions at the end points. In recent
proposals \cite{majoranawires}, these Majorana fermions are braided around each other, which
hopefully leads to the detection of non-abelian statistics. Kitaev's famous model
of interaction spin-1/2 particles on the honeycomb lattice \cite{k06}, has, in the presence
of a magnetic field, a gapped phase, whose excitations are of Ising type. Several ways of realizing
such a model have been proposed, see for instance \cite{zst07}. Finally, we mention a new
class of materials, the topological insulators. Fu and Kane \cite{fk08} proposed a system
of a 3D topological insulator, coated with an $s$-wave superconductor, resulting a
chiral $p$-wave superconductor, harboring Majorana bound states at its vortices.

To answer the question why one would be interested in correlation functions of an arbitrary number of $\sigma$ fields, we take an excursion to the quantum Hall effect. The quantum Hall effect, observed at filling fraction $\nu=\frac{5}{2}$ is attributed to the formation of the Moore-Read quantum Hall state \cite{mr91}. Theoretically, this quantum Hall state is `constructed' from the chiral Ising conformal field theory (combined with a compactified chiral boson). The operator creating an electron, consists of a Majorana fermion, combined with a vertex operator of the chiral boson, which is associated to the charge of the electron. The wave function obtained in this way, contains a pfaffian factor (coming from the Majorana fermions). The operator creating an excitation (or quasi-hole), contains the spin-field $\sigma$, apart from a vertex operator. The recent excitement about the $\nu = \frac{5}{2}$ quantum Hall effect originates in the properties of these $\sigma$ fields, which (hopefully) describe the excitations of this state. 

Most importantly, the $\sigma$ field has non-trivial fusion rules. That is, upon fusion two $\sigma$'s, there are two possible outcomes: $\sigma\times\sigma=\mathbf{1}+\psi$. The other fusion rules read
$\sigma \times \psi = \sigma$ and $\psi \times \psi = \mathbf{1}$. For a correlator to be non-zero, it is necessary that all field can be fused to the identity $\mathbf{1}$. In the case of four $\sigma$ fields, this can be done in two, independent ways. As a result, the correlator of four $\sigma$'s 'stands for' two conformal blocks. 

On the level of the quantum Hall effect, this has the following interpretation. Creating four excitations in the Moore-Read state (for instance, by increasing the magnetic field), can give two different results. These two different states differ in their topological properties. It has been proposed to use these different states as states of a qubit, which inherently is protected from decoherence by the environment, due to its topological nature \cite{k03,dfn05}.
Braiding the quasi-holes around each other, has the effect of acting by unitary matrices. These matrices have been shown not to commute, which underlies the nomenclature of `non-abelian' quantum Hall state.

To observe effects of non-abelian statistics, on has to do a measurement on the $\nu=\frac{5}{2}$ quantum Hall state. Most measurements in the quantum Hall effect involve some kind of (charge) transport measurement. In the quantum Hall effect, the charge is transported via the edge states, which are described with an (edge) conformal field theory. The charge and statistics of excitations can be probed by measuring the response of so called point contacts, which are constrictions, in which two edges are brought close together, allowing particles to tunnel from one edge to another. By measuring the shot-noise in a single point contact, the fractional charge of the excitations of the $\nu=\frac{1}{3}$ Laughlin state has been confirmed \cite{fraccharge}.
To get a hand on the statistics properties is harder, but attempts in that direction have been made in double point contact 'interferometers' \cite{doublepc}, trying, amongst other things, to
observe the predicted `even-odd effect' for the $\nu=5/2$ state \cite{topint}

To calculate the response of these constricted geometries, on uses the edge state formalism, expressing the response in term of conformal field theory correlators. In the non-abelian case, such a program has been carried out for single and double point contact
\cite{ffn07,dpc}.
In those calculations, one ends up with (to lowest order in perturbation theory) four point functions. These four point functions are know for most conformal field theories. However, in going to different geometries, or in higher order in perturbation theory, one will  encounter higher order correlators. For `free' theories, such as the compactified $u(1)$ chiral boson theories, these correlators are well known. As indicated above, it is extremely hard to find such correlators for arbitrary conformal field theories. The notable (and physically very relevant) exception is the Ising conformal field theory. In this paper, we will explicitly give the (chiral) correlators of an arbitrary (even) number of $\sigma$ fields, and an arbitrary number of $\psi$ fields.

Another use of the chiral correlators to which this paper is devoted in the context of quantum
Hall wave functions is the issue of the effect of braiding non-Abelian particles around each
other. It has been conjectured that, if one expresses the wave functions in terms of conformal
blocks, one only needs to consider the monodromies of the conformal blocks, while the
Berry phase does not give a contribution to the (non-Abelian) statistics \cite{mr91}.
For the abelian Laughlin state, this follows from an explicit Berry phase calculation
\cite{asw84}. Progress in proving this conjecture in the non-Abelian case was made in
\cite{bw92,nw96,gn97} and more recently in \cite{r09} and \cite{bgn10up}.
Having access to the explicit expressions presented in this paper might
facilitate such calculations in a more general setting%
\footnote{%
We were informed by the authors of \cite{bgn10up}, that they also obtained
the Ising correlators as we present here.
}.

The knowledge of the Ising chiral correlators has yet another application which is the 
 computation of the conformal blocks of the WZW model based
on  the $su(2)_2$ current algebra. These correlators will also be given in this paper 
using the fact that this WZW model
can be expressed as the product of the Ising model times a compactified boson. 
In the case of the $su(2)_1$ WZW model there is a unique chiral correlator
involving an arbitrary  number  of spin 1/2 primary fields. This correlator has a Jastrow type
form which,  interestingly enough,  gives the ground state of 
 the Haldane-Shastry Hamiltonian 
for  a spin 1/2 chain  with inverse square exchange interactions \cite{hs88,cs00}. 
In a similar way,  the $su(2)_2$ WZW chiral correlators can be used to construct the  ground states
of a non-abelian version of the  Haldane-Shastry model \cite{cs00b}. These results suggest  an interesting
analogy between spin systems and Fractional Quantum Hall systems having a common conformal
field theory underlying structure. 

We will close this introduction by going back to the early days of conformal field
theory. Shortly after the seminal paper of Belavin, Polyakov and Zamolodchikov \cite{bpz84},
Dotsenko and Fateev expressed the correlation functions of arbitrary minimal models,
in terms of contour integrals \cite{df84,df85}, based on the Coulomb gas formalism.
Despite these expressions, in many calculations it is advantageous
to use more explicit expressions. The connection between the integral formulation of
Dotsenko and Fateev and the expressions presented here is an interesting problem.

This paper is organized as follows. In section 2, we will review some basic  properties
of the Ising conformal field theory, and give some simple correlators as examples.
In section 3, we will show how one can obtain a simple explicit form of the correlator
of $2n$ $\sigma$ fields, by using a bosonized form of the conformal blocks.

Section 4 contains the results on the arbitrary conformal blocks of the Ising model.
These blocks are obtained in two different ways, giving rise to a curious set of
identities between bosonic and fermionic forms of these conformal blocks. In section
5, we present a different form of the conformal blocks which will be useful in future
applications. Finally, before we conclude, we use the results obtained in the earlier
sections, to present the arbitrary conformal blocks of the WZW theory based on the
$su(2)_2$ current algebra in section 6. 

\section{Some preliminaries on the Ising conformal field theory}

The critical Ising model is the simplest conformal field theory (CFT)
amongst the `minimal' model CFT's studied by Belavin, Polyakov and Zamolodchikov,
\cite{bpz84}. This theory has three primary fields $\id$, $\varepsilon$ and $\sigma$,
whose conformal dimensions $(h,\bar{h})$  are $(0,0)$, $(1/2,1/2)$ and $(1/16,1/16)$
respectively. The fusion rules read
\begin{align}
\label{fusionrules}
\nonumber
\id \times \id &= \id \\
\id \times \sigma &= \sigma & \sigma \times \sigma &= \id + \sigma \\
\nonumber
\id \times \varepsilon &= \varepsilon & \sigma \times \varepsilon &= \sigma &
\varepsilon \times \varepsilon = \id \ ,
\end{align}
or in terms of the fusion coefficients,
\begin{equation}
N_{\id x x} = N_{x \id x} = N_{\sigma \sigma \id} = N_{\sigma \sigma \varepsilon}
= N_{\sigma \varepsilon \sigma} = N_{\varepsilon \sigma \sigma} = N_{\varepsilon \varepsilon \varepsilon} \ ,
\end{equation}
where $x$ stands for any of the fields, and all the other coefficients are zero. 

These rules imply that the number of conformal blocks (CB) involved in a correlator of
an (even) number $2n$ primary fields $\sigma$ and $N$ primary fields
$\varepsilon$ is $2^{n-1}$.
The field $\varepsilon$ can be written as the product of a chiral Majorana fermion
$\psi(z)$ and an antichiral Majorana fermion $\bar{\psi} (\bar{z})$, i.e.
$\varepsilon (z,\bar{z}) = \psi(z) \bar{\psi} (\bar{z})$. The conformal blocks are the
building blocks of the non-chiral correlation functions (see below), and we will denote
them by
\begin{equation} 
\cF_{\bm}^{2n,N} (v_1,\ldots,v_{2n},z_1,\ldots,z_{N})
= \langle \sigma (v_1)\cdots \sigma (v_{2n})\psi (z_1)\cdots \psi(z_{N}) \rangle_{\bm} \ .
\end{equation}
Here, the vector $\bm$ of length $n$ labels the conformal block. The $i^{\rm th}$ entry
of $\bm$ specifies the fusion channel of the fields $\sigma(v_{2i-1})$ and $\sigma(v_{2i})$.
If $m_i=0$, they fuse to the trivial particle $\id$, while for $m_i=1$, they fuse to $\psi$.
Naively, this gives rise to $2^{n}$ different labels for the conformal blocks, but for
the CB to be non-zero, there have to be an even number of $m_i=1$. We will find that 
the explicit expressions we will give below, do in fact not depend on $m_1$, so
we indeed obtain the right number $2^{n-1}$ of different conformal blocks.

In this paper, we will give two
essentially different forms for the conformal blocks of an arbitrary number of $\sigma$
fields $2n$ and an arbitrary number $N$ of fields $\psi$. From these
conformal blocks, one can obtain the full, non-chiral correlation functions as follows
\begin{equation}
\label{corre}
\langle \sigma (v_1,\bar{v}_1)\cdots \sigma (v_{2n},\bar{v}_{2n})
\varepsilon (z_1,\bar{z}_1)\cdots \varepsilon (z_{N},\bar{z}_{N}) \rangle = 
\sum_\bm \cF_{\bm}^{2n,N} \overline{\cF}_{\bm}^{2n,N} \ ,
\end{equation}
where $\overline{\cF}_{\bm}^{2n,N}$ is the complex conjugate of
$\cF_{\bm}^{2n,N}$. In the remainder of the paper, we will mainly be
concerned with the chiral conformal blocks.

Belavin, Polyakov and Zamolodchikov derived a set of second order differential
equations, whose solutions are the conformal blocks $\cF_{\bm}^{2n,N}$,
which take the form
\begin{align}
\label{BPZ}
\Bigl( \frac{4}{3} \frac{\partial^2}{\partial v_a^2} -
\sum_{b\neq a} \frac{1}{(v_a-v_b)}\frac{\partial}{\partial v_b} -
\sum_{i} \frac{1}{(v_a-z_i)}\frac{\partial}{\partial z_i} -
\frac{1}{16}\sum_{b\neq a} \frac{1}{(v_a-v_b)^2} -
\frac{1}{2}\sum_{i} \frac{1}{(v_a-z_i)^2}
\Bigr) \cF_{\bm}^{2n,N} &= 0 \\
\Bigl( \frac{3}{4} \frac{\partial^2}{\partial z_i^2} -
\sum_{a} \frac{1}{(z_i-v_a)}\frac{\partial}{\partial z_i} -
\sum_{j\neq i} \frac{1}{(z_i-z_j)}\frac{\partial}{\partial z_j} -
\frac{1}{16}\sum_{a} \frac{1}{(z_i-v_a)^2} -
\frac{1}{2}\sum_{j\neq i} \frac{1}{(z_i-z_j)^2}
\Bigr) \cF_{\bm}^{2n,N} &= 0 \ .
\end{align}

We will frequently make use of the (chiral)
operator product expansion (OPE), which describes the behaviour of two
fields in the limit they approach each other, and reflect the fusion rules 
\eqref{fusionrules}. In particular, we have
\begin{align}
\sigma(z) \sigma(w) & \sim \frac{1}{(z-w)^{1/8}} + \frac{1}{\sqrt{2}} (z-w)^{3/8} \psi(w) &
\psi(z) \psi(w) &\sim \frac{1}{(z-w)} \ .\\
\end{align}
The factor $\frac{1}{\sqrt{2}}$ stems from the constant
$C_{\sigma\sigma\varepsilon} = \frac{1}{2}$
which appears in the OPE of the non-chiral fields
\begin{equation}
\sigma(z,\bar{z})\sigma(w,\bar{w}) 
\sim \frac{1}{|z-w|^{1/4}} + C_{\sigma\sigma\varepsilon} |z-w|^{3/4} \varepsilon(w,\bar{w}) \ .
\end{equation}
We note that these OPE coefficients can be calculated from (four point) correlators, which
are specified by the differential equations they satisfy. The constant $C_{\sigma\sigma\id}=1$
normalizes the fields $\sigma$.

Before we start with the correlator of an arbitrary (even) number of $\sigma$ fields, we
will first deal with the simpler two and four point correlators. The two point correlators
(as well as the three-point correlators) are determined by global conformal symmetry.
In particular we have
\begin{equation}
\cF^{2,0} = \langle \sigma(v_1) \sigma (v_2) \rangle = \frac{1}{(v_1-v_2)^{1/8}} = v_{12}^{-1/8} \ ,
\end{equation}
where we introduced the notation $v_{ab} = v_a - v_b$, and the exponent is given by
$-2h_\sigma$.

The four-point correlators are not completely determined by global conformal symmetry.
However, one can always make a transformation, which transforms the variables
$(v_1,v_2,v_3,v_4)$ to $(x,0,1,\infty)$, where $x=\frac{v_{12}v_{34}}{v_{14}v_{32}}$
is the cross-ratio. The partial differential equations for the four-point function transform
into an ordinary differential equation, which is in general much easier to solve.
Typically, this gives rise to hypergeometric functions, but in the case of the spin field of the
Ising CFT, the result is simpler:
\begin{equation}
\label{foursigma}
\cF^{4,0}_{\bm} =\langle \sigma (v_1)\sigma (v_2)\sigma (v_3)\sigma (v_4) \rangle_{\bm} = 
\frac{1}{\sqrt{2}} (v_1-v_2)^{-\frac{1}{8}}(v_3-v_4)^{-\frac{1}{8}}
\sqrt{(1-x)^{\frac{1}{4}}+(-1)^{m_2} (1-x)^{-\frac{1}{4}}} \ ,
\end{equation}
$\bm$ labels the two conformal blocks, the possible values being $(0,0)$ and
$(1,1)$, for pairwise fusion of the $\sigma$ fields to the $\id$ and $\psi$ channel
respectively. Note that the expression does not depend on $m_1$. The overall
factor is fixed by the requirement that in the limit $v_2\rightarrow v_1,v_4\rightarrow v_3$,
one obtains, by making use of the OPE, the value $1$ for the correlator in the case
$\bm=(0,0)$. The same limit gives rise to the correlator
$\langle \psi(v_1)\psi(v_3)\rangle = \frac{1}{v_1-v_3}$ in the case $\bm=(1,1)$.

\section{The $2n$ point $\sigma$ correlator}
\label{sigmacor}

In this section, we will present a particular simple form of the correlator of an
arbitrary even number of $\sigma$ fields. This will allow us to use the operator
product expansion to obtain expressions for conformal blocks of an arbitrary
(even) number of $\sigma$ fields and an arbitrary number of $\psi$ fields.
We will obtain these correlators in both a `bosonic' and `fermionic' form, giving rise to
a curious set of functional identities, some of which appear to be new.
It is well known that one can bosonize the Ising model \cite{fsz87}, which allows
one to obtain the correlation functions. By starting from the results of \cite{fsz87},
Fendley, Fisher and Nayak \cite{ffn07}, provide a systematic way of writing down
the chiral $2n$-point correlation functions of $\sigma$-fields, in the representation
in which the $\sigma$ fields are pairwise in a definite fusion channel, in terms of
a bosonized correlator, which we will give below. We will take this expression as
our starting point to obtain the results presented in this paper. 

To be able to write the CB's in a compact way, 
we define the following cross ratios, first we pair the variables $(v_{2i-1},v_{2i})$,
and for each pair of pairs, we introduce
\begin{equation}
x_{i,j} = \frac{(v_{2i-1}-v_{2i})(v_{2j-1}-v_{2j})}{(v_{2i-1}-v_{2j})(v_{2j-1}-v_{2i})} \ ,
\end{equation}
for $i,j=1,\ldots,n$ and the $v_i$ are the locations of the $\sigma$ fields. 
So, for $4$ fields, we have one cross-ratio, consistent with global conformal transformations.
What might be a little surprising is that we introduce, for $2n$ large enough, more
cross-ratios than there are variables, namely $n (n-1)/2$ of them. Even though in general,
not all these cross-ratios are independent, the expressions for the correlators are rather
simple in terms of this over complete set of cross-ratios.

Before we explicitly give the form of the $2n$ point correlators, we should point out again that
there are $2^{n-1}$ fusion channels. We will label these fusion channels by
a vector with $n$ entries, namely $\bm=(m_1,m_2,m_3,\ldots,m_{n})$, there the
$m_i$ take the values $0$ or $1$. The expression for the $2n$ point $\sigma$ correlator
we present below does not depend on the value of $m_1$. Thus, we indeed have the
correct number of independent conformal blocks. We will start by giving the expression
of the $2n$-point chiral $\sigma$ correlator as given by Fendley, Fisher and Nayak in
terms of the chiral correlator of a compactified boson field $\phi$.
\begin{equation}
\label{bosscor}
\bigl( \langle \sigma(v_1)\sigma(v_2) \cdots \sigma(v_{2n}) \rangle_{\bm} \bigr)^2 \propto
\biggl\langle \prod_{j=1}^{n} \Bigl( 
e^{i(\phi(v_{2j-1})-\phi(v_{2j}))/2} +
(-1)^{m_j} e^{-i(\phi(v_{2j-1})-\phi(v_{2j}))/2}
\Bigr)
\biggr\rangle
\end{equation}
The correlator of the vertex operators of the chiral boson $\phi$ can easily be
evaluated, by making use of the result
\begin{equation}
\langle e^{i a_1 \phi} (v_1) \cdots e^{i a_n \phi} (v_n) \rangle = \prod_{i<j} (v_i-v_j)^{a_i a_j} \ ,
\quad {\rm when}  \qquad \sum_i a_i = 0 \ .  
\end{equation}
This correlator vanishes when $\sum_i a_i \neq 0$. Swapping the signs of all the $a_i$
does not change the correlation function. 

By using this result \eqref{bosscor} of \cite{ffn07},
we obtain the following expression for the $2n$ point correlator
\begin{equation}
\label{scor}
\langle \sigma(v_1)\sigma(v_2) \cdots \sigma(v_{2n}) \rangle_{\bm} =
2^{-\frac{1}{2}(n-1)}
\prod_{i=1}^{n} (v_{2 i-1}-v_{2i})^{-\frac{1}{8}}
\sqrt{\sum_{\substack{t_1=1\\t_2,t_3,\ldots,t_{n}=-1,1}}
\prod_{i=1}^{n} t_i^{m_i} \prod_{1\leq i<j\leq n} (1-x_{i,j})^{\frac{t_i t_j}{4}}} \ ,
\end{equation}
where the normalization is fixed by requiring that for $\bm =(0,0,\ldots,0)$, the correlator
reduces to $1$ if one fuses all pairs $(2i-1,2i)$ of $\sigma$-fields ($i=1,\ldots,n$). The
correlator $\langle \sigma(v_1) \sigma(v_2)\rangle = (v_1-v_2)^{-\frac{1}{8}}$
gives the normalization of the $\sigma$-field. Also, note that the correlator \eqref{scor} does
not depend on the value of $m_1$, which reduces the number of independent correlators
ot $2^{n-1}$

Let us first use the above expression with $\bm = (1,1,\ldots,1)$ to reduce
a correlator of $2n$ $\sigma$-fields (with $2n$ a multiple of four)
to a correlator of $n$ $\psi$-fields, which is just the Pfaffian.
In doing so, we will make use of the operator product expansion (OPE)
\begin{equation}
\label{ope}
\lim_{v_2\rightarrow v_1} \sigma (v_1) \sigma (v_2) =
(v_1-v_2)^{-\frac{1}{8}}\id + (v_1-v_2)^{\frac{3}{8}} C_{\sigma,\sigma}^{\psi} \sigma (v_1)
+ \text {higer order terms} \ ,
\end{equation}
with the (chiral) OPE coefficient $C_{\sigma,\sigma}^{\psi} = \frac{1}{\sqrt{2}}$

Using this OPE several times, by taking the appropriate
limits, namely $v_{2i} \rightarrow v_{2i-1}$, for $i=1,\ldots,n$
we find that the correlator eq. \eqref{scor} with $\bm = (1,1,\ldots,1)$ reduces to
(after a relabeling of the
variables $v_{2i-1}\rightarrow z_i$, with $i=1,\ldots,n$, and $N=n$)
\begin{equation}
\label{psifromsigma}
\langle \psi (z_1) \psi (z_2) \cdots \psi (z_{N}) \rangle =
\sqrt{\hf \Bigl( \frac{1}{(z_i-z_j)^2} \Bigr)} \ , 
\end{equation}
where $\hf (M)$ denotes the haffnian of a symmetric $N \times N$ matrix $M$,
and is given by
$\hf (M) = \frac{1}{2^{N/2}(N/2)!}\sum_{\sigma \in \cS_{N}}
\prod_{i=1}^{N/2}M_{\sigma(2i-1),\sigma(2i)}$.
In other words, the haffnian of a matrix $M_{ij}$ is obtained by summing over all different ways of
pairing the indices, where each term contributes a factor
$M_{i_{1}i_{2}} M_{i_{3}i_{4}}\cdots M_{i_{N-1}i_{N}}$.

To convince oneself that eq. \eqref{psifromsigma} follows from eq. \eqref{scor} by making use
of the OPE we note that
\begin{equation}
\begin{split}
&\lim_{v_{2i}\rightarrow v_{2i-1}} (C_{\sigma,\sigma}^{\psi})^{-n}
\prod_{i=1}^{n} (v_{2 i-1}-v_{2i})^{-\frac{3}{8}}
\langle \sigma(v_1)\sigma(v_2) \cdots \sigma(v_{2n}) \rangle_{(1,1,\ldots,1)} =
\\&
\lim_{v_{2i}\rightarrow v_{2i-1}}
\prod_{i=1}^{n} (v_{2 i-1}-v_{2i})^{-\frac{1}{2}}
\sqrt{\sum_{\substack{t_1=1\\t_2,t_3,\ldots,t_{n}=-1,1}}
\prod_{i=1}^{n} t_i \prod_{1\leq i<j\leq n} (1-{\frac{t_i t_j}{4}}x_{i,j} )} \ ,
\end{split}
\end{equation}
because in the limit $v_{2i}\rightarrow v_{2i-1}$, $x_{i,j}\rightarrow 0$.
Now, in the expansion of $\prod (1-t_i t_j x_{i,j}/4)$, only those terms wich contain all indices on $x$ once (and only once) survive in the limit. If an index $i$ appears more than once, the term will vanish, because
we get an overall contribution proportional to at least $(v_{2i-1}-v_{2i})^{1/2}$, which vanishes in the limit. If an index $i$ does not appear in the expansion of
$\prod (1-t_i t_j x_{i,j}/4)$, then there will be a similar term in the sum over the $t_i$'s, but with the opposite overall sign (namely, if the index $i\neq 1$, it is the term in which $t_i$ has the opposite value; for $i=1$, it is the term in which all $t_j$ have the opposite value). Thus, we conclude that only those terms, in which each index appears once and only once will be present. It is not hard to convince oneself that all these terms have positive sign (because the $t_i$ will enter as $t_i^2$).
In the terms present, the numerator of the $x_{i,j}$ present in the expansion will be cancelled by the prefactor $\prod (v_{2 i-1}-v_{2i})^{-\frac{1}{2}}$ (thus making the limit well defined). The denominators which remain, give rise to the haffnian. In the limit, the denominators of the term
$x_{i,j} x_{k,l} \ldots$ give rise to factors
$1/(v_{2i-1}-v_{2j-1})^2 1/(v_{2k-1}-v_{2l-1})^2 \cdots$.
Because  in the expansion of
$\prod (1-t_i t_j x_{i,j}/4)$ only terms in which all indices appear once and only once are present, we end up with a (unsigned) sum over all possible ways of picking pairs $v_{2i-1},v_{2j-1}$, and for each pair, we have a factor $1/(v_{2i-1}-v_{2j-1})^2$. This is precisely the haffnian
$\hf \Bigl( \frac{1}{(v_{2i-1}-v_{2j-1})^2} \Bigr)$. It is not hard to check that the overall factors of $2$ work out as well. The OPE coefficients
$C_{\sigma,\sigma}^{\psi}=1/\sqrt{2}$ give rise to a factor $2^{\frac{n}{2}}$. The factors of $1/4$ in products contribute $2^{-\frac{n}{2}}$ (after taking the square root). Finally, the sum over the $t_i$ gives a factor of
$2^{\frac{n-1}{2}}$. Combining all this with the prefactor $2^{-\frac{n-1}{2}}$, we indeed find that the factors of $2$ cancel. 

To show that the reduction described above gives rise to the pfaffian
$\pf \Bigl( \frac{1}{(z_i-z_j)} \Bigr)$,
we note a famous identity between the pfaffian and haffnian, namely
\begin{equation}
\label{pfhf}
\Biggl[ \pf \Bigl( \frac{1}{(z_i-z_j)} \Bigr)\Biggr]^2 = \hf \Bigl( \frac{1}{(z_i-z_j)^2} \Bigr) \ , 
\end{equation} 
where $\pf (A)$ denotes the pfaffnian of an anti-symmetric $N\times N$ matrix $A$,
and is given by 
$\pf (A) = \frac{1}{2^{N/2}(N/2)!}\sum_{\sigma \in \cS_{N}}
\prod_{i=1}^{N/2} \sign(\sigma) A_{\sigma(2i-1),\sigma(2i)}$.
This identity follows from the Cauchy identity.

\section{Correlators of an arbitrary number of $\sigma$'s and $\psi$'s.}

We will now continue with the description of the conformal blocks of both $\sigma$ and $\psi$ fields.
Such correlators were known for two \cite{mr91} and four \cite{nw96} $\sigma$ fields, and an arbitrary number of $\psi$ fields. Because we have an explicit result for the correlator of an arbitrary (even) number of $\sigma$ fields, we can also obtain the mixed correlators of an arbitrary number of $\psi$ fields and an arbitrary even number of $\sigma$ fields, by making use of the OPE. We will derive these results explicitly for two $\sigma$ fields and an arbitrary number of $\psi$ fields. For the other cases, we will merely state the (somewhat cumbersom) results.

Apart from the method sketched above, there is an other way to obtain the arbitrary conformal
blocks of the Ising model, which uses the approach of \cite{nw96} (see \cite{as07} for a
generalization of the results of \cite{nw96} to $\mathbf{Z}_k$ parafermions and parafermions
based on $su(3)_2$). In this approach, one constructs the blocks by starting from the $\psi$
correlators, with the correct (polynomial) dependence of the coordinates of the $\sigma$ particles
build in. There are $2^{n-1}$ such functions which are linearly independent, and one
considers a general linear combination,
where the `coefficients' will depend on the coordinates of the $\sigma$ fields (we note that
a `manifestly' independent set is given in reference \cite{rr96}).
These `coefficients'
are obtained by performing appropriate fusions, to obtain already known correlators. In this way,
one constructs a set of $2^{n-1}$ independent functions, which have the right degree, zero
and pole structure. This constitutes a proof that the functions obtained are indeed the
conformal blocks of the Ising model. We refer to \cite{nw96,as07} for more details on this
approach.

Thus, we can obtain different, but necessarily  
equivalent, expressions for the correlators, generalizing the Haffnian-Pfaffian identity \eqref{pfhf}.
We will refer to the expressions obtained by using the bosonization procedure (i.e. those
containing the `Haffnians') as the bosonic form, while we refer to the expressions obtained
by the procedure outlined in the previous paragraph (i.e. those which will contain the
`Pfaffians') as the fermionic form.

In the context of the Moore-Read state, this identity for two $\sigma$ fields already noted in \cite{mr91},
while for more $\sigma$'s, these identities appear to be knew.

\subsection{The case of two $\sigma$ fields.}
The case of two $\sigma$ fields and an arbitrary number of $\psi$ fields is the last
example we will deal with in some detail, the results for the remaining cases will
be simply stated. The arguments are completely equivalent, however. 

We consider a correlator of $2+2N$ $\sigma$-fields, with $\bm = (0,1,1,\ldots,1)$.
We fuse all but the first two fields in pairs. The result is
\begin{equation}
\label{scred}
\langle \sigma(v_1) \sigma(v_2) \psi(z_1) \cdots \psi(z_N) \rangle = 
2^{-\frac{N}{2}} (v_1-v_2)^{-\frac{1}{8}}
\sqrt{\sum_{I} 2^{|I|} \Bigl(\hf_{i,j\in I}\frac{1}{(z_i-z_j)^2}\Bigr)
\frac{(v_1-v_2)^{|\widetilde{I}|}}{\prod_{j\in \widetilde{I}}(v_1-z_j)(v_2-z_j)}} \ ,
\end{equation}
where $N$ is even and the sum is over all subsets  of $\{1,2,\ldots,N\}$, containing an even
number of elements $|I|$. The set $\widetilde{I}$ (containing $|\widetilde{I}|$ elements) is equal to
$\{1,2,\ldots,N\}\backslash I$.

We will know briefly describe how this result can be obtained. Even though we will not take the limit
$v_2 \rightarrow v_1$, we still have that all $x_{i,j}\rightarrow 0$ in the limit
$v_{2i} \rightarrow v_{2i-1}$, for $i=2,\ldots,N+1$. As before, we have the result that in the expansion of
$\prod (1-t_i t_j x_{i,j}/4)$, any index $i>1$ must appear once and only once. This implies we will have an even number $r$ of factors $t_i x_{1,i}$, which in the limit give rise to a factor
$(v_1-v_2)^r/\prod_{i}(v_1-v_{2i-1})(v_2-v_{2i-1})$. The remaining factors $t_{j} t_{j'} x_{j,j'}$ give rise to a haffnian in the `remaining' variables, in the same way as in the previous section. 
Let us check the factors of two. Both the OPE coefficients as well as the sum over the $t_i$ give a factor of $2^{N/2}$. The number of factors of $x$ depends on $r$, which results in an overall factor of
$2^{(-N-r)/2}$. Combined with the two factors $2^{N/2}$, this gives the factor $2^{|I|}$ in the square root. The prefactor $2^{-N/2}$ originates in the prefactor in \eqref{scor}. 

Due to an identity already noted in \cite{mr91}, namely
\begin{equation}
\label{pfiden}
\sum_{I} 2^{|I|} (v_1-v_2)^{N-|I|}\Bigl(\hf_{i,j\in I}\frac{1}{(z_i-z_j)^2}\Bigr)
\prod_{j\in I}(z_j-v_1)(z_j-v_2) = \Bigl[
 \pf \Bigl( \frac{(z_i-v_1)(z_j-v_2)+(z_i-v_2)(z_j-v_1)}{z_i-z_j} \Bigr) \Bigr]^2\ ,
\end{equation}
equation \eqref{scred} reduces to the result which normally appears in the literature,
namely
\begin{equation}
\begin{split}
&\langle \sigma(v_1) \sigma(v_2) \psi(z_1) \cdots \psi(z_N) \rangle =
2^{-\frac{N}{2}} (v_1-v_2)^{-\frac{1}{8}}
\times \\ & \times
\prod_{j=1}^{N} (v_1-z_j)^{-\frac{1}{2}}(v_2-z_j)^{-\frac{1}{2}}
 \pf \Bigl( \frac{(z_i-v_1)(z_j-v_2)+(z_i-v_2)(z_j-v_1)}{z_i-z_j} \Bigr) \ .
\end{split}
\end{equation}

We can repeat the above exercise for $2 + 2N$, with $N$ an odd integer. This means
that the first two $\sigma$-fields now also should fuse to a $\psi$. It turns out that in
this case, the correlator takes exactly the form \eqref{scred}, but now with $N$ being
an odd integer. Note that in this case, all the terms in the square root are proportional
to $(v_1-v_2)$, which is not the case for $N$ even.
It is not that hard to generalize the relation \eqref{pfiden} to an odd number of $z$
variables as well, namely for $N$ an odd integer, we have
\begin{equation}
\begin{split}
\label{pfidenodd}
&\sum_{I} 2^{|I|} (v_1-v_2)^{N-|I|-1}\Bigl(\hf_{i,j\in I}\frac{1}{(z_i-z_j)^2}\Bigr)
\prod_{j\in I}(z_j-v_1)(z_j-v_2) = \\ &
\Bigl[\sum_{m=1}^{N} (-1)^m
 \pf_{i,j\neq m} \Bigl( \frac{(z_i-v_1)(z_j-v_2)+(z_i-v_2)(z_j-v_1)}{z_i-z_j} \Bigr) \Bigr]^2\ .
\end{split}
\end{equation}

\subsection{The case of four $\sigma$ fields.}
Before we give the most general result, we will first check if we can reproduce the result obtained in
\cite{nw96}, for the correlator of four $\sigma$-fields, and an arbitrary even
number of $\psi$-fields. Using contractions of equation \eqref{scor}, we find the following, somewhat involved, expression
\begin{multline}
\label{foursigpsi}
\langle \sigma(v_1)\sigma(v_2)\sigma(v_3)\sigma(v_4)\psi(z_1)\cdots\psi(z_N)\rangle_{\bm} =
2^{-\frac{N+1}{2}}(v_1-v_2)^{-\frac{1}{8}}(v_3-v_4)^{-\frac{1}{8}}\times \\
\sqrt{
\sum_{I} \biggl[
2^{|I|}\Bigl(\hf_{i,j\in I}\frac{1}{(z_i-z_j)^2}\Bigr)
\Biggl(
\sum_{\widetilde{I}_1,\widetilde{I}_2}
\Bigl( \frac{(-1)^{|\widetilde{I}_2|+m_2}}{(1-x)^{\frac{1}{4}}}+(1-x)^{\frac{1}{4}} \Bigr)
\frac{(v_1-v_2)^{|\widetilde{I}_1|}}{\prod_{i_1\in \widetilde{I}_1}(v_1-z_{i_1})(v_2-z_{i_1})}
\frac{(v_3-v_4)^{|\widetilde{I}_2|}}{\prod_{i_2\in \widetilde{I}_2}(v_3-z_{i_2})(v_4-z_{i_2})}
\Biggr)
\biggr]
} \ .
\end{multline}
Some remarks are in order here. As above, the sum over $I$ is over all subsets of $\{1,\ldots,N\}$ with an even number of elements, while sum over the sets $\widetilde{I}_1$ and
$\widetilde{I}_2$ is over all possible ways to divide the set $\{1,\ldots,N\}\backslash I$
into two sets, whose order is important. 

In \cite{nw96}, the case of four $\sigma$'s and an arbitrary (but even) number of $\psi$'s
was  obtained differently. The expression obtained makes use of the following
functions
\begin{equation}
\Psi_{(k_1 k_2)(k_3 k_4)} = \pf\Bigl(
\frac{(v_{k_1}-z_i)(v_{k_2}-z_i)(v_{k_3}-z_j)(v_{k_4}-z_j)+
(v_{k_1}-z_j)(v_{k_2}-z_j)(v_{k_3}-z_i)(v_{k_4}-z_i)}{z_i-z_j} \Bigr) 
\end{equation}
The three functions $\Psi_{(12)(34)}$, $\Psi_{(13)(24)}$ and $\Psi_{(14)(23)}$ are not independent, but satisfy
$\Psi_{(14)(23)} = x \Psi_{(12)(34)}+(1-x) \Psi_{(13)(24)}$.
In terms of these functions, we can write the correlator \eqref{foursigpsi}
in the following form
\begin{multline}
\label{foursigpsi2}
\langle \sigma(v_1)\sigma(v_2)\sigma(v_3)\sigma(v_4)\psi(z_1)\cdots\psi(z_N)\rangle_{\bm} =
2^{-\frac{N+1}{2}}(v_1-v_2)^{-\frac{1}{8}}(v_3-v_4)^{-\frac{1}{8}}
\prod_{i=1}^{4}\prod_{j=1}^{N} (v_i-z_j)^{-\frac{1}{2}} \times\\
\times
\Bigl((1-x)^{\frac{1}{4}}+\frac{(-1)^{m_2}}{(1-x)^{\frac{1}{4}}}\Bigr)^{-\frac{1}{2}}
\bigl((1-x)^{\frac{1}{4}}\Psi_{(13)(24)} +(-1)^{m_2}(1-x)^{-\frac{1}{4}}\Psi_{(14)(23)}
\bigr) \ .
\end{multline}
We should point out that \eqref{foursigpsi} and \eqref{foursigpsi2} differ by at most
a sign.

\subsection{The case of an arbitrary even number of $\sigma$ fields}

We will now provide an expression for the conformal blocks of $2n$ $\sigma$ fields
and $N$ $\psi$ fields, with both $N$ and $n$ integers.
We first give the expression, and explain the notation afterwards.
\begin{multline}
\label{nsignpsi}
\langle \sigma(v_1)\cdots\sigma(v_{2n})\psi(z_1)\cdots\psi(z_N)\rangle_{\bm} =
2^{-\frac{N+n-\delta_{n>0}}{2}}\prod_{i=1}^{n}(v_{2 i-1}-v_{2 i})^{-\frac{1}{8}}\times \\ \times
\sqrt{
\sum_{I} 
2^{|I|}\Bigl(\hf_{i,j\in I}\frac{1}{(z_i-z_j)^2}\Bigr)
\biggl[
\sum_{\widetilde{I}_1,\ldots,\widetilde{I}_{n}} 
A^{(\bm+\br) \bmod 2}_{2n} \prod_{i=1}^{n}
\frac{(v_{2i-1}-v_{2i})^{r_i}}{\prod_{j\in \widetilde{I}_i}(v_{2i-1}-z_{j})(v_{2i}-z_{j})}
\biggr]
} \ .
\end{multline}
In the formula above, we used the following notation. First of all, the sum over $I$ is over
all subsets of $\{1,\ldots,N\}$ with an even number of elements. The sum over
$\widetilde{I}_1,\ldots,\widetilde{I}_{n}$ is over all different ways of dividing
the set $\{1,\ldots,N\}\backslash I$ into $n$ subsets, which might be empty, and the order of these
sets is important. Furthermore, $r_i=|\widetilde{I}_i|$ is the number of elements of
$\widetilde{I}_i$ and $\br=(r_1,\ldots,r_{n})$. We use the convention that
$\sum_i m_i = N \bmod 2$, and the vector $(\bm + \br) \bmod 2$ has elements
$((m_1+r_1)\bmod2,\ldots,(m_{n}+r_{n})\bmod2)$. 
The function $A^{\bm}_{2n}$ is closely related to the correlator of $2n$ $\sigma$ fields
\begin{equation}
A^{\bm}_{2n} = \sum_{\substack{t_1=1\\t_2,t_3,\ldots,t_{n}=-1,1}}
\prod_{i=1}^{n} t_i^{m_i} \prod_{1\leq i,j \leq n} (1-x_{i,j})^{\frac{t_i t_j}{4}} \ .
\end{equation}
Finally, we introduced the notation $\delta_{n>0}$, which is $1$ if $n>0$, and zero for
$n=0$.

It is not to hard to generalize equation \eqref{foursigpsi2} to an arbitrary (but for now, even)
number of $\psi$ fields. To do so, we start with the `preferred basis' for the functions
$\Psi$, as described in \cite{nw96}. Nayak and Wilczek label the $2^{n-1}$ functions
with two $n$ tuples, which have the property that $2i-1$ and $2i$ are never in the same
tuple. Note that the order of the two tuples is irrelevant. For us, it will be convenient to
label the functions $\Psi$ by the vector $\bt=(t_1,t_2,\ldots,t_{n})$. The relation with
the two tuples is a follows. If $t_i=1$, we have $2i-1$ in the first tuple and $2i$ in the
second, while for $t_i=-1$ the situation is reversed. As an example, $\Psi_{(1,-1,1)}$
corresponds to $\Psi_{(145)(236)}$. We will write the `preferred basis' functions $\Psi$
explicitly in terms of the $t_i$
\begin{equation}
\label{psit}
\Psi_{\bt} = \pf \biggl(
\frac{\prod_{i=1}^{n} (v_{(2i-1)-\frac{t_i-1}{2}}-z_j)(v_{(2i)+\frac{t_i-1}{2}}-z_k)
+\prod_{i=1}^{n}(v_{(2i-1)-\frac{t_i-1}{2}}-z_k)(v_{(2i)+\frac{t_i-1}{2}}-z_j)}{(z_j-z_k)} \biggr) \ .
\end{equation}
The subscripts on the $v$'s is a little involved, but for $t_i=1$, the pair
$(v_{(2i-1)-\frac{t_i-1}{2}},v_{(2i)+\frac{t_i-1}{2}})$ is simply $(v_{2i-1},v_{2i})$, 
while for $t_i=-1$, we get $(v_{2i},v_{2i-1})$. 
We note that it is in principle possible to write $\Psi$ related to an arbitrary pair of
tuples in terms of the functions \eqref{psit}, generalizing the relation
$\Psi_{(12)(34)} = -\frac{1-x}{x} \Psi_{(13)(24)}+\frac{1}{x} \Psi_{(14)(23)}$.

We have now introduced all the notation necessary to
write down the generalization of \eqref{foursigpsi2} to an arbitrary even number of
$\sigma$ and an even number of $\psi$ fields
\begin{multline}
\label{nsignpsi2}
\langle \sigma(v_1)\cdots\sigma(v_{2n})\psi(z_1)\cdots\psi(z_N)\rangle_{\bm} = \\
2^{-\frac{N+n-\delta_{n>0}}{2}}\prod_{i=1}^{n}(v_{2i-1}-v_{2i})^{-\frac{1}{8}}
\prod_{i=1}^{2n}\prod_{j=1}^{N} (v_i-z_j)^{-\frac{1}{2}} 
\bigl(A_{2n}^{\bm}\bigr)^{-\frac{1}{2}}
\biggl(\sum_{\substack{t_1=1\\t_2,t_3,\ldots,t_{n}=-1,1}} 
\Bigl(\prod_{i=1}^{n} t_i^{m_i} \prod_{1\leq i,j \leq n} (1-x_{i,j})^{\frac{t_i t_j}{4}}\Bigr) 
\Psi_{\bt}
\biggr) \ .
\end{multline}
To generalize \eqref{nsignpsi2} to an odd number of $\psi$ fields, we just have to
perform a contraction of (say) the first $\psi$ field and one of the $\sigma$ fields. For convenience, we choose to fuse the first $\psi$ at $z_1$ with the first $\sigma$, at $v_1$. It turns out that we only have to modify the form of $\Psi_\bt$ to obtain the generalization of \eqref{nsignpsi2} to an odd number of $\psi$ fields. Namely, we define, for $N$ odd,
\begin{multline}
\label{psitodd}
\Psi_{\bt} =  (v_1-v_2)^{\frac{1}{2}} 
\sum_{m=1}^{N} \biggl[(-1)^m \biggl(\prod_{i=2}^{n}
\frac{(v_1-v_{2i+\frac{t_i-1}{2}})^\frac{1}{2}}{(v_1-v_{2i-1-\frac{t_i-1}{2}})^{\frac{1}{2}}} 
(v_{2i-1-\frac{t_i-1}{2}}-z_m) \biggr)\times \\ \times
\pf_{j,k\neq m} \biggl(
\frac{\prod_{i=1}^{n} (v_{(2i-1)-\frac{t_i-1}{2}}-z_j)(v_{(2i)+\frac{t_i-1}{2}}-z_k)
+\prod_{i=1}^{n}(v_{(2i-1)-\frac{t_i-1}{2}}-z_k)(v_{(2i)+\frac{t_i-1}{2}}-z_j)}{(z_j-z_k)} \biggr)
\biggr] \ .
\end{multline}
By using this form for $\Psi_\bt$ in \eqref{nsignpsi2} when $N$ is odd, we obtain a expression
for the correlator of an even number of $\sigma$ fields, and an odd number of $\psi$ fields.
We could have fused the $\psi$ field with any other $\sigma$ field as well. This would have led to different, but equivalent, correlators.
For instance, if we fused the $\psi$ at $z_1$ to the $\sigma$ at $v_2$, instead of $v_1$ as we did above, we would need to do the following replacement in eq. \eqref{psitodd}
\begin{equation}
\biggl(\prod_{i=2}^{n}
\frac{(v_1-v_{2i+\frac{t_i-1}{2}})^\frac{1}{2}}{(v_1-v_{2i-1-\frac{t_i-1}{2}})^{\frac{1}{2}}} 
(v_{2i-1-\frac{t_i-1}{2}}-z_m)
\biggr)
\rightarrow
\biggl(\prod_{i=2}^{n}
\frac{(v_2-v_{2i-1-\frac{t_i-1}{2}})^\frac{1}{2}}{(v_2-v_{2i+\frac{t_i-1}{2}})^{\frac{1}{2}}} 
(v_{2i+\frac{t_i-1}{2}}-z_m) \biggr) \ .
\end{equation}
If we would have fused $z_1$ to any other $v_j$, we would have obtained slightly more complicated expressions (basically, because our expressions do not depend on $m_1$). We will not
give these expressions here, but we trust that the interested reader can work them out.

The fact that the arbitrary chiral Ising correlators can be obtained in two independent ways, gives rise to what appear to be new identities. For completeness, we will give this identity in its most general form. This identity is based on the equivalence between the correlators in eq. \eqref{nsignpsi} and eq.
\eqref{nsignpsi2}. Namely, we have
\begin{equation}
\begin{split}
\label{bosfermid}
&\sum_{I} 
2^{|I|}\Bigl(\hf_{i,j\in I}\frac{1}{(z_i-z_j)^2}\Bigr)
\biggl[
\sideset{}{'}\sum_{\widetilde{I}_1,\ldots,\widetilde{I}_{n}} 
A^{(\bm+\br) \bmod 2}_{2n} \prod_{i=1}^{n}
\frac{(v_{2i-1}-v_{2i})^{r_i}}{\prod_{j\in \widetilde{I}_i}(v_{2i-1}-z_{j})(v_{2i}-z_{j})}
\biggr] = \\ &
\prod_{i=1}^{2n}\prod_{j=1}^{N} (v_i-z_j)^{-1} 
\bigl(A_{2n}^{\bm}\bigr)^{-1}
\biggl(\sum_{\substack{t_1=1\\t_2,t_3,\ldots,t_{n}=-1,1}} 
\Bigl(\prod_{i=1}^{n} t_i^{m_i} \prod_{1\leq i,j \leq n} (1-x_{i,j})^{\frac{t_i t_j}{4}}\Bigr) 
\Psi_{\bt}
\biggr)^2
\end{split}
\end{equation}
Here, we used the same notation as above, and $\Psi_{\bt}$ is given by eq. \eqref{psit} for $N$ even, while it is given by eq. \eqref{psitodd} for $N$ odd.

\section{Alternative representation}

In this section, we provide an alternative representation of the fermionic form of
the conformal blocks of an arbitrary number of $\sigma$ and $\psi$ fields,
which will be useful later on. To obtain the representation we present in this
section,  one only has to reshuffle some overall factors.
In particular, we will consider the form where there is an overall factor
$\prod_{1\leq a < b \leq n} (v_a-v_b)^{-1/8}$, while the factor of the type
$\prod_{i,a}(z_i - v_a)^{-1/2}$ will be completely absorbed in the `Pfaffian part'
of the wave function.

\subsection{Conformal blocks of $\sigma$ fields}

The simplest correlator for which there is a difference between the two representations is the
4-$\sigma$ correlator, for which the two possible CB's can be written as 
\beq
  {\cal F}^{4}_{\bf 0,1}= \langle \s(v_1) \,  \s(v_2) \,  \s(v_3) \,  \s(v_3)\rangle_{\bf 0,1}  = 2^{-1/2}
  \prod_{a < b } v_{a b} ^{-1/8} \left( \sqrt{ v_{13} \, v_{24}}  \pm \sqrt{v_{14} \,  v_{23} } \right)^{1/2} 
\label{CB2}
\eeq
where the block ${\bf p}=0$ (resp. ${\bf p}= 1$) corresponds to the plus sign (resp. minus sign)
on the RHS. 

Let us write eq. (\ref{CB2}) in the following form 

\beq
  {\cal F}^{4}_{\bf p}= 2^{-1/2}
  \prod_{a < b } v_{a b} ^{-1/8} \left(  \ep_{{\bf p} {\bf  0}} \sqrt{ v_{13} \,  v_{24}}  + \ep_{{\bf p} {\bf  1}}
    \sqrt{v_{14} \, v_{23} } \, \right)^{1/2} 
\label{CB2b}
\eeq
where $\ep_{{\bf p} {\bf  q}}$ are sign factors and ${\bf  q}$ label
the monomials in the summand on the RHS. The values of these quantities are
\beq
\begin{array}{cl}
\ep_{{\bf 0} {\bf  0}} = 1, & \ep_{{\bf 0} {\bf  1}} = 1 \\
\ep_{{\bf 1} {\bf  0}} = 1,  & \ep_{{\bf 1} {\bf  1}} = -1 \\
\end{array}
\label{epsi2}
\eeq

The expression of the CB's for 6 $\sigma$-fields has a structure
similar to eq.(\ref{CB2}),

\beq
  {\cal F}^{6}_{\bf p}= 2^{-1}
  \prod_{a < b } v_{a b} ^{-1/8} \left(  \ep_{{\bf p} {\bf  0}} \sqrt{ v_{135} \,  v_{246}}  + \ep_{{\bf p} {\bf  1}}  \sqrt{v_{136} \, v_{245} }  +  \ep_{{\bf p} {\bf  2}} \sqrt{ v_{146} \, v_{235}}  + \ep_{{\bf p} {\bf  3}}  \sqrt{v_{145}\,  v_{236} }  
  \,  \right)^{1/2} 
\label{CB3}
\eeq
where 
\beq
v_{a bc} = v_{ab} \, v_{ac} \,  v_{bc}
\label{vvv}
\eeq
The four possible CB's correspond to the following choices
\beq
\begin{array}{clll}
\ep_{{\bf 0} {\bf  0}} = 1, & \ep_{{\bf 0} {\bf  1}} = 1,  & \ep_{{\bf 0} {\bf  2}} = 1, & \ep_{{\bf 0} {\bf  3}} = 1 \\
\ep_{{\bf 1} {\bf  0}} = 1, & \ep_{{\bf 1} {\bf  1}} = -1,  & \ep_{{\bf 1} {\bf  2}} = 1, & \ep_{{\bf 1} {\bf  3}} = -1 \\
\ep_{{\bf 2} {\bf  0}} = 1, & \ep_{{\bf 2} {\bf  1}} = 1,  & \ep_{{\bf 2} {\bf  2}} =- 1, & \ep_{{\bf 2} {\bf  3}} = -1 \\
\ep_{{\bf 3} {\bf  0}} = 1, & \ep_{{\bf 3} {\bf  1}} = -1,  & \ep_{{\bf 3} {\bf  2}} =- 1, & \ep_{{\bf 3} {\bf  3}} = 1 \\
\end{array}
\label{epsi3}
\eeq
One can verify for these choices that (\ref{CB3}) satisfies the BPZ eq.(\ref{BPZ}), and that
they correspond to the expression given in section \ref{sigmacor}.

\begin{figure}[ht]
\begin{center}
\includegraphics[height= 3.5 cm]{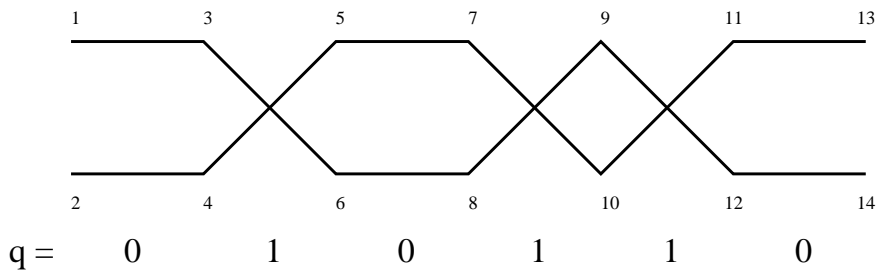}
\end{center}
\caption{Illustration of the macrogroups, corresponding to
$(1,3,6,8,9,12,14)(2,4,5,7,10,11,13)$. The value for $q=22$, or in binary digits, $q = 010110$.}
\label{macrogroup}
\end{figure}

Consider a CB of  $2 n$ $\s$-fields whose coordinates are $v_1, \dots, v_{2 n}$, 
and pair them as $(v_1, v_2) (v_3, v_4) \dots (v_{2 n -1}, v_n)$, which we will call
reference pairs. We might as well
use the labels of these coordinates so that the reference pairs are given by
$(1,2)(3,4) \dots (2n-1,2n)$. 
A basis of CB's is associated to two {\em macrogroups} containing $n$ elements each
\beq
 ({\ell_1},  {\ell_2},  \dots, {\ell_n}), \qquad (  {\ell'_1}, {\ell'_2}, \dots, {\ell'_n}) \ .
\label{macro}
\eeq
such that two members of a reference pair never belong to the same macrogroup. 
As an example, take the case $n=3$. The four possible pairs of macrogroups are given by
$(135)(246), (136)(245), (146)(235), (145)(236)$. For $n$ integer,  there are $2^{n-1}$
different macrogroup pairs  (\ref{macro}),  
which can be labelled by an integer ${\bf q}=0, 1, \dots, 2^{n-1}-1$.  
The macrogroup $ \vec{\ell}= ({\ell_1},  {\ell_2},  \dots, {\ell_n})$
 associated to  ${\bf q}$ is given by

\beq
\begin{array}{ll}
\ell_1 = 1 & \\
\ell_{k+1} - \ell_k = 2 & {\rm if} \; q_k = 0 \\
\ell_{k+1} - \ell_k = 1 & {\rm if} \; q_k = 1 \; \; {\rm and} \; \ell_k: {\rm even} \\
\ell_{k+1} - \ell_k = 3 & {\rm if} \; q_k = 1 \; \; {\rm and} \; \ell_k: {\rm odd} \\
\end{array}  
\label{recur} 
\eeq
where $q_k$ are the binary digits of the integer ${\bf q} = (q_1, q_2, \dots, q_{n-1})$. 
 The macrogroup  $\vec{\ell'} = (  {\ell'_1}, {\ell'_2}, \dots, {\ell'_n})$  associated to ${\bf q}$ also satisfies
 the recursion relations   (\ref{recur})  plus the initial condition  $\ell'_1 = 2$. The  macrogroup pair
associated to ${\bf q} =0$ are
$(1,3,\dots, 2n -1)(2,4, \dots, 2n)$. A graphical representation of this construction
is given in  terms of a two leg ladder where the $k^{\rm th}$-rung  $(k=1, \dots, n)$ corresponds to the reference pair
$(2k-1,2k)$. The pair $(\vec{\ell}_{\bf q} , \vec{\ell'}_{\bf q})$ is described by two strings of integers on the ladder, which 
on the $k^{\rm th}$-plaquette crosses or not  for 
 $q_k=1$ or  $q_k = 0$ respectively  (see fig.  \ref{macrogroup}).  
 As explained  earlier,  a macrogroup
 is associated to a set of $n$ coordinates $v_{\ell_k}$. We shall now assign an overall
 factor to it, 
\beq
v_{\vec{\ell}} = \prod_{k < m }^n  v_{\ell_k, \ell_m}, \qquad
v_{\ell_k, \ell_m} = v_{\ell_k} - v_{\ell_m},
\label{vmacro}
\eeq
and similarly for the macrogroup $\vec{\ell' }$.  For $n=1$ we take by convenium 
$v_{\vec{\ell}} = v_{\vec{\ell'}}=1$. For $n=2$,  we get $v_{(\ell_1, \ell_2)}= v_{\ell_1} - v_{\ell_2}$, 
which agrees with the earlier definition of $v_{ab} = v_a - v_b $.  The case $n=3$ 
coincides with the earlier definition of $v_{abc}$ given in eq.(\ref{vvv}) which
corresponds to the macrogroup $(abc)$.

The connection between the $q_i$ and the $t_i$ can be described as follows. First, we
consider $(1-t_i)/2$, which takes the values $0$ and $1$ for $t_i=1$ and $t_i=-1$ respectively.
Then, the $q_i$ are given by $q_i = \sum_{j=1}^{i+1} (1-t_{j})/2 \bmod 2$. 

\begin{figure}[ht]
\begin{center}
\includegraphics[height= 3.5 cm]{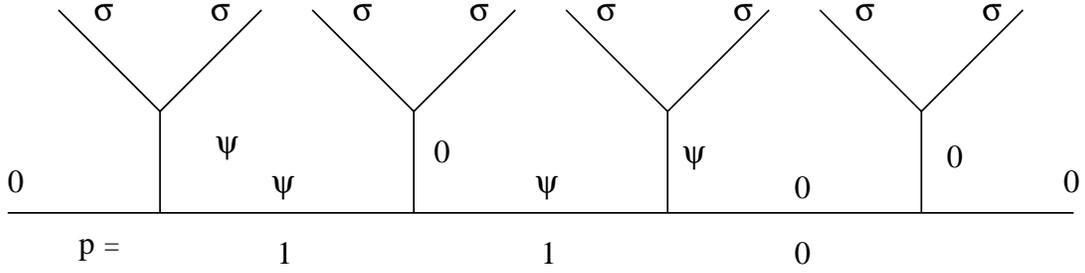}
\end{center}
\caption{Label of a correlator, specifying the conformal block. For the block displayed, $\bp = 6$,
or in binary notation, $\bp = 110$.}
\label{conformal}
\end{figure}

The CB's can be labeled by graphs whose links are associated
to primary fields which meet at vertices whenever the coincident fields satisfy the fusion rules. 
The primary fields appearing in the correlator are associated to the external lines
of the graph, while the primary fields of the internal lines gives rise to the existence
of different CB's. This picture is the familiar one in particle physics if we view
the primary fields as particles and the CB's as scattering processes.  Fig \ref{conformal} shows
 the scattering representation of the  CB (\ref{CBf0}), according to which the fields $\sigma(v_{2k-1})$
 and $\sigma(v_{2 k})$ fuse together giving rise to either the identity or the Majorana field. These fields
 in turn fuse together on a base line. The binary digits of ${\bf p}$ 
 describe the primary fields running on this base line: the identity if $p_i=0$ or 
 the Majorana field if $p_i=1$.

The connection with the labels of the CB's in the previous sections is as follows. There, the
labels $m_i$ indicated the fusion channel of the pair $(v_{2i-1},v_{2i})$, with $i=1,\ldots n$.
$m_i=0$ for the fusion into the identity channel, while $m_i=1$ for the $\psi$ channel. It
follows that the relation with the $p_i$ reads $p_i = \sum_{j=1}^{i} m_j \bmod 2$. 

After these definitions we can finally give
the expression of the CB of $2n$ $\s$-fields and no  Majorana fermions, 
\beq
{\cal F}^{2 n, 0}_{\bf p} = C_{n,0} \prod_{a<b}^{2 n} v_{ab}^{-1/8} \; 
\left(  \sum_{{\bf q}=0}^{2^{n-1}  -1} \ep_{{\bf p} {\bf q}} \sqrt{  v_{\ell_{\bf q}}   v_{\ell'_{\bf q}}  }
\right)^{1/2}, \qquad {\bf p} =0, 1, \dots , 2^{n-1}-1, 
\label{CBf0}
\eeq
where $\ep_{{\bf p} {\bf q}}$ is a sign given in terms of the binary digits of ${\bf p}$ and $ {\bf q}$ 
\beq
\ep_{{\bf p} {\bf q}} = (-1)^{\sum_{k=1}^{n-1} p_k q_k} 
\label{epq}
\eeq
and the constant $C_{n,0}$ is given by
\beq
C_{n,0} = 2^{-(n-1)/2} \ .
\label{constantC}
\eeq
The latter result can be checked taking the limit $v_{2k-1} \rightarrow v_{2k}$ and using the
OPE eq. (\ref{ope}). 

 As explained in eq.(\ref{corre}), the CB can be used to construct the non chiral correlator
 of fields. For the Ising model these correlators are known thanks to a bosonic version
 of two copies of the Ising model. In particular one has
 
  \beq
 \langle \sigma(v_1, \bar{v}_1) \dots \sigma(v_{2 n}, \bar{v}_{2 n})  \rangle^2  = 
 2^{-n} \sum_{
 \epsilon_i = \pm 1,  \; 
  \sum  \epsilon_i =0}   \; \;  \prod_{i < j} \, |v_i - v_j|^{ \epsilon_i \epsilon_j/2} 
 \label{corre2}
 \eeq

We have checked for $n=2,3$ that 

\beq
 \langle \sigma(v_1, \bar{v}_1) \dots \sigma(v_{2 n}, \bar{v}_{2 n})  \rangle^2  = 
 \left( 
 \sum_{\bf p}
   {\cal F}^{2 n, 0}_{\bf p}\,   {\cal \bar{F}}^{2 n, 0}_{\bf p} \right)^2 \ .
\eeq

\subsection{Conformal blocks with $\sigma$ and $\psi$ fields}

The chiral correlator or CB of an even number $2 m$ of Majorana fields is given by 
\beq
{\cal F}^{(0,2m)}(z_1, \dots, z_{2 m}) = 
\langle \psi(z_1) \dots \psi(z_{2 m}) \rangle = {\rm Pf} \bigl(\frac{1}{z_i - z_j}\bigr) , \qquad 
\label{majo}
\eeq
where ${\rm Pf}$ is the Pfaffian of the $2 m  \times 2 m$ antisymmetric matrix $1/(z_i - z_j)$.
For a generic antisymmetric matrix $A_{i j}$ the Pfaffian is given by
$$
{\rm Pf} \, A = \sqrt{ {\rm det} A} =  \frac{1}{m! \, 2^m} \sum_{ \pi \in {\cal S}_{2 m} } {\rm sgn} (\pi) 
\prod_{i=1}^n A_{\pi(2 i -1), \pi(2i)} \ ,
$$
where ${\cal S}_{2 m}$ is the permutation group of $2 m$ symbols. For an odd
number of Majorana fields the CB is zero. 

Let us next consider the CB of two $\sigma$ fields and an even number of $\psi$ fields. 
There is only one CB, whose expression was found by  Moore and Read \cite{mr91}

$$
\langle \s(v_1) \,  \s(v_2)  \, \prod_{i= 1}^{2 m} \chi(z_i) \rangle  =  2^{-m}  \; 
v_{12}^{-1/8} \,  \prod_{i=1}^{2 m}   \left( (z_i-  v_1) ( z_i-v_2) \right)^{-1/2} 
\; {\rm Pf} \;  \bigl(  \frac{ (z_i - v_1)( z_j - v_2) + (z_i - v_2) (z_j - v_1)}{z_i - z_j} \bigr)  \ .
$$
For later purposes it is convenient to write this eq. as
\beq
{\cal F}^{(2,2m)}_{\bf 0} (v_1, v_2, z_1, \dots, z_{2 m}) = 
  2^{-m}  \;  v_{12}^{-1/8} \, {\rm Pf} \; \Bigl( \frac{ h_{(1),(2)}(z_i, z_j)}{z_i-z_j}\Bigr)
 \label{F22m} 
 \eeq
where
\beq
h_{(1),(2)}(z_i, z_j) =  \left[ \frac{ (z_i-v_1)(z_j - v_2)}{(z_i - v_2)(z_j-v_1)} \right]^{1/2} +
( i \leftrightarrow j) \ .
\label{h12}
\eeq

For four $\sigma$ fields and $2m$ $\psi$ fields there are two CB's which were
found by Nayak and Wilczek, which can be written as \cite{nw96} 
\barray 
{\cal F}^{(4,2m)}_{\bf 0,1} (v_1,\dots,  v_4, z_1, \dots, z_{2 m}) & =  & 
  C   \;   \prod_{a < b} v_{ab}^{-1/8} \, \left(   \sqrt{ v_{13} v_{24}}  
 \pm   \sqrt{ v_{14} v_{23}}  \right)^{-1/2}  \label{F42m} \\
 & \times &  \left[  \sqrt{ v_{13} v_{24}}  \; {\rm Pf} \; \Bigl(\frac{ h_{(13),(24)}(z_i, z_j)}{z_i-z_j}\Bigr) 
 \pm   \sqrt{ v_{14} v_{23}} \;   {\rm Pf} \; \Bigl( \frac{ h_{(14),(23)}(z_i, z_j)}{z_i-z_j}\Bigr) \right] 
\nonumber 
 \earray
where
\beq
h_{( a b),( c d)}(z_i, z_j) =  \left[ \frac{ (z_i-v_a)(z_i - v_b) (z_j-v_c)(z_j - v_d)}{(z_i - v_c)(z_i-v_d)(z_j-v_a)(z_j - v_b)} \right]^{1/2} +
( i \leftrightarrow j) \ .
\label{h123}
\eeq
Notice the analogy between equations (\ref{F42m})  and (\ref{CB2}). The subindices of the $h$ matrix elements
are nothing but the two macrogroups $(13)(24)$ and $(14)(23)$ associated to the CB of four $\sigma$ fields.
Indeed, one can write the following generalization of eqs. (\ref{F22m}, \ref{F42m}) 
\barray 
{\cal F}^{(2 n ,2m)}_{\bf p} (v_1, \dots,  v_{2 n}, z_1, \dots, z_{2 m}) & =  & 
C_{2 n, 2m}  \prod_{a<b}^{2 n} v_{ab}^{-1/8} \; 
\left(  \sum_{{\bf q}=0}^{2^{n-1}  -1} \ep_{{\bf p} {\bf q}} \sqrt{  v_{\ell_{\bf q}}   v_{\ell'_{\bf q}}  }
\right)^{-1/2} \label{F2n2m} \times \\
& \times & 
\left[   \sum_{{\bf q}=0}^{2^{n-1}  -1} \ep_{{\bf p} {\bf q}} \sqrt{  v_{\ell_{\bf q}}   v_{\ell'_{\bf q}} }
\; \; {\rm Pf} \; \frac{ h_{ \ell_{\bf q}, \ell'_{\bf q}}(z_i, z_j)}{z_i - z_j}
\right]
\nonumber 
\earray 
where
\beq
h_{\vec{\ell}, \vec{\ell'}}(z_i, z_j) =  \left[ 
\prod_{k=1}^n \frac{ (z_i-v_{\ell_k})(z_j - v_{\ell'_k}) }{ (z_i-v_{\ell'_k})(z_j - v_{\ell_k})} \right]^{1/2} +
( i \leftrightarrow j)
\label{hgeneral}
\eeq
and
$$
C_{2 n, 2m} = 
\left\{
\begin{array}{ll} 
1 & n=0 \\
2^{ - ((n-1)/2 + m)} & n >0 \ .\\
\end{array} 
\right. 
$$
It is not hard to convince oneself that this expression is equivalent to eqs.
\eqref{nsignpsi2} and \eqref{psit}.
We also checked for small values of $n$ and $m$ that (\ref{F2n2m}) satisfies the
BPZ equations (\ref{BPZ}). 

Let us know consider the CB with an odd number of fermions.
For $m>1$ we start from the equation 
\beq
{\cal F}^{(2,2m)}_{\bf 0} (v_1, v_2, z_1, \dots, z_{2 m}) = 
  2^{-m}  \;  v_{12}^{-1/8} \, {\rm Pf} \; \Bigl(\frac{ h_{(1),(2)}(z_i, z_j)}{z_i-z_j}\Bigr)
 \eeq
and take the limit $z_{2 m} \rightarrow v_1$. One gets
$$
\frac{ h_{(1),(2)}(z_i, z_{2 m})}{z_i-z_{2m} }
\rightarrow \frac{v_{12}^{1/2}}{ (z_{2m} - v_1)^{1/2}} [(z_i - v_1) (z_i - v_2)]^{-1/2}
$$
Next we use the expansion of the Pfaffian of a $2m \times  2m$ matrix 
$$
{\rm Pf} \, A = \sum_{i=1}^{2 m-1} (-1)^{1+i} \, A_{i, 2m} \; {\rm Pf}_{j,k \neq i, 2m} \; A_{j, k}
$$ 
where the RHS contains the Pfaffian of the matrix obtaining from $A$ by deleting the $i^{\rm th}$
and $(2 m)^{\rm th}$ rows and columns. Using again the OPE of $\psi(z_{2m})$ and $\sigma(v_1)$
one gets
\beq
{\cal F}^{(2,2m-1)}_{\bf 0} (v_1, v_2, z_1, \dots, z_{2 m-1}) = 
  2^{-m+ 1/2}  \;  v_{12}^{3/8} \, \sum_{i=1}^{2m -1} (-1)^{ 1+i} \,  [(z_i - v_1) (z_i - v_2)]^{-1/2} \; 
{\rm Pf}_{j,k \neq i, 2m} \; \Bigl( \frac{ h_{(1),(2)}(z_j, z_k)}{z_j-z_k} \Bigr) \ .
 \label{F22m_1} 
 \eeq
For generic CB's involving an even number of $\sigma$ fields one has 
in the limit $z_{2 m} \rightarrow v_1$
$$
\frac{ h_{\vec{\ell}, \vec{\ell}' }(z_i, z_{2 m})}{z_i-z_{2m} }
\rightarrow \frac{(z_i - v_1)^{-1}}{ (z_{2m} - v_1)^{1/2}} \frac{ \prod_{k=1}^n \sqrt{v_{1 \ell'_k}}}{ \prod_{k=2}^n \sqrt{v_{1 \ell_k}}}
\prod_{k=1}^n \left( 
\frac{ z_i - v_{\ell_k} }{ z_i - v_{\ell'_k}} \right)^{1/2} 
$$
which gives 
\barray 
{\cal F}^{(2 n ,2m-1)}_{\bf p} & =  & 
C_{2 n, 2m-1}  \prod_{a<b}^{2 n} v_{ab}^{-1/8} \; 
\left(  \sum_{{\bf q}=0}^{2^{n-1}  -1} \ep_{{\bf p} {\bf q}} \sqrt{  v_{\ell_{\bf q}}   v_{\ell'_{\bf q}}  }
\right)^{-1/2} 
\\
& \times & 
\left[   \sum_{{\bf q}=0}^{2^{n-1}  -1} \ep_{{\bf p} {\bf q}} \sqrt{  v_{\ell_{\bf q}}   v_{\ell'_{\bf q}} }
\; \;     \frac{ \prod_{k=1}^n \sqrt{v_{1 \ell'_k}}}{ \prod_{k=2}^n \sqrt{v_{1 \ell_k}}}  \right. \nonumber \\
& \times  & \left. \sum_{i=1}^{2 m -1} (-1)^{i+1} 
 \; (z_i - v_1)^{-1} \, 
\prod_{k=1}^n \left( 
\frac{ z_i - v_{\ell_k} }{ z_i - v_{\ell'_k}} \right)^{1/2} 
 \; \; 
{\rm Pf}_{j,k \neq i, 2m}\; \Bigl(\frac{ h_{ \ell_{\bf q}, \ell'_{\bf q}}(z_j, z_k)}{z_j - z_k}\Bigr)
\right]
\nonumber 
\earray 
with 

$$
C_{2 n, 2m-1} = 
\left\{
\begin{array}{ll} 
1 & n=0 \\
2^{ - (n/2 + m-1)} & n >0 \ .\\
\end{array}
\right. 
$$
Again, one can check that this expression is equivalent to eqs. \eqref{nsignpsi2} and
\eqref{psitodd}.

\section{Free field representation of $su(2)_k$ for $k=1,2$}

In this section, we will use the expressions we found for the chiral Ising
correlators, to give the correlators of the primary fields of the WZW conformal
field theory based on $su(2)_2$. For more detail on the correlators and
conformal blocks of primary fields in the WZW model, we refer to the seminal
paper by Knizhnik and Zamolodchikov \cite{kz84}. The free field representation
for WZW theories was introduced by Wakimoto \cite{w86}
(for more information, we refer to, f.i. \cite{bmp90}).

Let us use the Cartan-Weyl basis for the generators of the $su(2)_k$
algebra $J^0(z), \, J^\pm(z)$. The OPE's are given by 
\barray
J^0(z) \; J^0(w) & \sim & \frac{ k/2}{ (z-w)^2} \nonumber \\
J^0(z) \; J^\pm(w) & \sim & \frac{ \pm J^\pm(w)}{ z-w} \nonumber \\
J^+(z) \; J^-(w) & \sim & \frac{ k}{ (z-w)^2}+ \frac{ 2  J^0(w)}{ z-w} \nonumber 
\earray 
For $k=1$ these currents can be realized as
$$
J^0(z) = \frac{i}{\sqrt{2}}  \,  \partial_z \, \phi(z), \qquad
J^\pm(z) =  e^{ \pm i \sqrt{2}  \phi(z)}
$$
The primary fields with spin $j= 1/2$ are
$$
V_{1/2,s}(z)  = e^{ i s \phi(z)/ \sqrt{2}}, \qquad s = \pm 1,  \qquad h_{1/2} = \frac{1}{4}
$$
For $k=2$ the currents can be written as
$$
J^0(z) = i \,  \partial_z \, \phi(z), \qquad
J^\pm(z) = \sqrt{2} \; \psi(z) \, e^{ \pm i \phi(z)}
$$
where $\psi(z)$ is the Majorana field of the Ising model. 
The primary fields with  spin $j= 1$ are
$$
V_{1,\pm 1}(z)  = e^{ \pm  i  \phi(z)},  \qquad  V_{1, 0}(z) = \psi(z) , \qquad 
  \qquad h_{1} = \frac{1}{2}
$$
while the primary fields with spin $j= 1/2$ are
$$
V_{1/2, s }(z)  =  \sigma(z) \,  e^{   i  s  \phi(z)/2} \qquad s = \pm 1,
  \qquad h_{1/2} = \frac{3}{16}
$$
where $\sigma(z)$ is the spin field of the Ising model, with $h_\sigma = 1/16$.

\subsection{Conformal blocks of $N$ spin $1$ fields}

Consider $N$ spins 1 labelled by $s_i = \pm 1, 0$ associated to the coordinates $z_i \; (i=1, \dots, N)$.
The wave function is given by
$$
\psi(s_1, \dots, s_N) = \chi_s  \; \prod_{i < j}^N  ( z_i - z_j)^{ s_i s_j} \; {\rm Pf}_0  \left(  \frac{1}{z_i - z_j} \right) 
$$
where
$$
\chi_s = (-1)^{ \sum_{i:even} (s_i -1)}
$$
and the Paffian is restricted to the positions where $s_i =0$.

Notice that the term 
$\prod_{i < j} ( z_i - z_j)^{ s_i s_j}$
only depends on the sites where $s_i = \pm 1$.
The total spin must be zero so
$\sum_{i=1}^N s_ i = 0$. Before we continue with the case combining the
spin $1$ and spin $1/2$ fields, we remark the following. For $N$ odd and
greater than one, one can form singlet(s) out of the spins. In principle, one could
expect a contribution from the case $s_i = 0$ for all $i$, which would correspond to
the correlator of an odd number of $\psi$ fields. However this correlator 
 is zero since it involves an odd number of $\psi$ fields. 

\subsection{Conformal blocks of $N$ spin $1$ and $2n$ spin $1/2$ fields}

Let $v_a \; (a=1, \dots, 2n)$ be the coordinates of the spin $1/2$ fields, $\tau_a= \pm 1$ 
their spin, and $z_i \;(i=1, \dots, N)$ the coordinates of the spin 1 fields, and $s_i=0, \pm 1$
their spin. We shall denote by $z_i^0$ the coordinates where $s_i = 0$, and assume there
are $M$ such coordinates. The total spin $S_z=0$ imposes that
$$
\frac{1}{2} \sum_{a=1}^{2 n}  \tau_{a}     + \sum_{i=1}^{N}  s_ i = 0
$$
The wave functions (or conformal blocks) are given, in terms of the
chiral Ising correlators, by  
\barray 
\psi_{\bf p}(\tau_1, \dots, \tau_{2n}, s_1, \dots, s_{N})  & = &  
\chi_\tau \chi_s  \; \prod_{a < b} v_{ab}^{  \tau_a \tau_b/4}    
\prod_{i < j}  z_{ij}^{ s_i s_j} \;  \prod_{i, a}  ( z_i - v_a)^{
\frac{1}{2} s_i \tau_a} \;  {\cal F}^{(2 n, M)}_{\bf p} (v_j, z_k^{(0)}) 
\label{psip}
\earray
where ${\cal F}^{(2 n, M)}_{\bf p} (v_j, z_k^{(0)}) $ is the Ising CB computed earlier and
the overall sign is given by
$$
\chi_s = (-1)^{ \sum_{i:even} (s_i -1)}, \qquad \chi_\tau = (-1)^{ \sum_{a: even} (\tau_a -1)/2} \ . 
$$

\section{Conclusion and Outlook}

In this paper, we gave explicit expressions for all the conformal blocks of the Ising model.
These expressions were obtained by starting from the bosonized formulation of the
Ising model, which gave us the expression for the conformal block of an arbitrary number
of $\sigma$ fields. From this, two different expressions for the correlators of an arbitrary
number of $\sigma$ and $\psi$ fields were obtained. The equivalence between these
expressions gave rise to a family of identities, which are a consequence of the fact that
the Ising model can be bosonized. We used the expressions for the conformal blocks of
the Ising model, to obtain expressions for the conformal blocks of the $su(2)_2$ WZW
conformal field theory.

It is straightforward to generalize these results to the conformal blocks of those WZW
theories, which can be written in terms of a set of chiral bosons, augmented with the
Ising conformal field theory. Theories of this type are $so(2n+1)_1$ and $E_8$ at
level $2$.

We hope that the explicit form of the Ising conformal blocks and correlators
presented in this paper, will aid in the design of experiments (and perhaps
applications) to detect the effects of non-abelian statistics, which has been
conjectured to be present in several different condensed matter systems,
ranging from the quantum Hall effect, $(p+ip)$ superconductors, systems
based on topological insulators, to cold atomic gases. 
In addition, the form of the correlators might inspire to make progress in
obtaining the explicit form of (typically much harder) multi-point correlators
in other conformal field theories.
 
\noindent
{\it Acknowledgments -} We are grateful to I. Cirac, for enlightning discussions
and to P.~Bonderson, V.~Gurarie and C.~Nayak for discussions on their paper \cite{bgn10up}
and comments on this manuscript.  
This work has been supported by the spanish project FIS2009-11654 (GS). 
We also acknowledge ESF Science Programme INSTANS 2005-2010 and
thank the Aspen Center for Physics (EA).

\end{document}